\documentclass[showpacs,10pt,twocolumn,prb]{revtex4-1}

\usepackage{amsmath}
\usepackage{amssymb}
\usepackage{graphicx}
\usepackage{amssymb}
\usepackage{graphics}
\usepackage{epsfig}
\usepackage{CJK}
\usepackage{color}

\begin{document}

\begin{CJK*}{GBK}{Song}
\title{Critical behavior of the van der Waals bonded ferromagnet Fe$_{3-x}$GeTe$_2$}
\author{Yu Liu ,$^{1}$ V. N. Ivanovski,$^{2}$ and C. Petrovic$^{1}$}
\affiliation{$^{1}$Condensed Matter Physics and Materials Science Department, Brookhaven National Laboratory, Upton, New York 11973, USA\\
$^{2}$Institute of Nuclear Sciences Vinca, University of Belgrade, Belgrade 11001, Serbia}
\date{\today}

\begin{abstract}
The critical properties of the single-crystalline van der Waals bonded ferromagnet Fe$_{3-x}$GeTe$_2$ were investigated by bulk dc magnetization around the paramagnetic (PM) to ferromagnetic (FM) phase transition. The Fe$_{3-x}$GeTe$_2$ single crystals grown by self-flux method with Fe deficiency $x \approx 0.36$ exhibit bulk FM ordering below $T_c = 152$ K. The M\"{o}ssbauer spectroscopy was used to provide information on defects and local atomic environment in such crystals. Critical exponents $\beta = 0.372(4)$ with a critical temperature $T_c = 151.25(5)$ K and $\gamma = 1.265(15)$ with $T_c = 151.17(12)$ K are obtained by the Kouvel-Fisher method whereas $\delta = 4.50(1)$ is obtained by a critical isotherm analysis at $T_c = 151$ K. These critical exponents obey the Widom scaling relation $\delta = 1+\gamma/\beta$, indicating self-consistency of the obtained values. With these critical exponents the isotherm $M(H)$ curves below and above the critical temperatures collapse into two independent universal branches, obeying the single scaling equation $m = f_\pm(h)$, where $m$ and $h$ are renormalized magnetization and field, respectively. The exponents determined in this study are close to those calculated from the results of the renormalization group approach for a heuristic model of three-dimensional Heisenberg ($d = 3, n = 3$) spins coupled with the attractive long-range interactions between spins that decay as $J(r)\approx r^{-(3+\sigma)}$ with $\sigma=1.89$.
\end{abstract}

\maketitle
\end{CJK*}

\section{INTRODUCTION}

Two-dimensional (2D) materials such as graphene and ultrathin transition-metal dichalcogenides exhibit a number of attractive properties that have been extensively studied in the past decade.\cite{Geim, Chhowalla, Hu, Bhimanapati} However, in contrast to mechanical and optoelectronic properties, the possibility of magnetism in 2D materials has received little attention. Recently, the van der Waals (VDW) bonded magnetic materials are of great interest as building blocks for heterostructures designed for application in spin-based information technologies. The CrX$_3$ (X = Cl, Br, I) and CrXTe$_3$ (X = Si, Ge, Sn) have been identified as promising candidate for long-range magnetism in monolayers.\cite{Zhang1, McGuire, Sivadas, Zhuang1, Lin1} CrSiTe$_3$ exhibits ferromagnetic (FM) ordering at 32 K in the bulk,\cite{Casto} and it can be enhanced to $\sim$ 80 K in monolayer and few-layer samples.\cite{Lin2} Bulk CrI$_3$ and CrGeTe$_3$ are ferromagnetic at $\sim$ 61 K, which is still somewhat low for spintronic applications.\cite{McGuire, Zhang2} Considering this, the VDW bonded material Fe$_{3-x}$GeTe$_2$ may be of particular interest because the bulk is ferromagnetic at $\sim$ 230 K.\cite{Deiseroth}

The ternary compound Fe$_{3-x}$GeTe$_2$ contains Fe$_{3-x}$Ge slabs with two inequivalent Fe sites Fe1 and Fe2.\cite{Deiseroth} The slabs are sandwiched between two VDW bonded Te layers, as depicted in Fig. 1(a). Fe$_{3-x}$GeTe$_2$ is a strongly correlated ferromagnetic metal with the Curie temperature $T_c \sim 230$ K or 220 K,\cite{Deiseroth, Zhu, Chen1} and the ferromagnetism with itinerant character can be tuned by controlling Fe content.\cite{Andrew} $T_c$ decreases with increasing Fe vacancies, and the lattice responds with a decrease in the in-plane lattice parameter and a slight expansion along the $c$ axis.\cite{Andrew} The flux-grown crystals typically have a lower $T_c \approx 150$ K with the Fe vacancies level $x \approx 0.3$.\cite{Andrew} The X-ray diffraction (XRD) and M\"{o}ssbauer spectroscopy reveal that the presence of Fe vacancies only occur in the Fe2 sites whereas no Fe atoms occupy the interlayer space; neutron powder diffraction (NPD) shows that the ratio of moments between Fe1 and Fe2 is 1.25 at 1.5 K in Fe$_{2.9}$GeTe$_2$ polycrystal.\cite{Yu} However, May $et$ $al.$ found that there is no significant difference in the moments on the two Fe atomic positions in the unit cell of flux-grown Fe-deficient single crystals and verified that the moments lie along the $c$ axis without any significant spin canting or reorientation.\cite{Andrew} Recently, in addition to the reported FM transition at 214 K in CVT-grown single crystals, Yi $et$ $al.$ determined that the ferromagnetic layers of Fe$_{2.9}$GeTe$_2$ actually order antiferromagnetically along the $c$ axis below 152 K.\cite{Yi} Furthermore, the density-functional calculations predict that single-layer Fe$_3$GeTe$_2$ is dynamically stable and exhibits a significant uniaxial magnetocrystalline anistropy, potentially useful for magnetic storage applications.\cite{Zhuang2}

In order to understand the magnetic behavior in few-layer samples and the possible applications of this material, it is necessary to establish the nature of magnetism in the bulk Fe$_{3-x}$GeTe$_2$. In this paper, we investigate the critical behavior of flux-grown Fe$_{3-x}$GeTe$_2$ single crystal by various techniques, such as a modified Arrott plot, Kouvel-Fisher plot, and critical isotherm analysis. Our analyses indicate that the obtained critical exponents [$\beta = 0.372\pm0.004$ ($T_c = 151.25\pm0.05$ K), $\gamma = 1.265\pm0.015$ ($T_c = 151.2\pm0.2$ K), and $\delta = 4.50\pm0.01$ ($T_c = 151$ K)] are close to those calculated from the results of the renormalization group approach for a three-dimensional Heisenberg spins coupled with a long-range interaction between spins decaying as $J(r)\approx r^{-(3+\sigma)}$ with $\sigma=1.89$.

\section{EXPERIMENTAL DETAILS}

High quality Fe$_{3-x}$GeTe$_2$ single crystals were grown by the self-flux technique starting from an intimate mixture of pure elements Fe (99.99 $\%$, Alfa Aesar) powder, Ge (99.999 $\%$, Alfa Aesar) pieces, and Te (99.9999 $\%$, Alfa Aesar) pieces with a molar ratio of 2 : 1 : 4. The starting materials were sealed in an evacuated quartz tube, which was heated to 1000 $^\circ$C over 20 h, held at 1000 $^\circ$C for 3 h, and then slowly cooled to 680 $^\circ$C at a rate of 1 $^\circ$C/h. X-ray diffraction (XRD) data were taken with Cu $K_{\alpha}$ ($\lambda=0.15418$ nm) radiation of a Rigaku Miniflex powder diffractometer. The element analysis was performed using energy-dispersive x-ray spectroscopy (EDX) in a JEOL LSM-6500 scanning electron microscope. The magnetization was measured in a Quantum Design magnetic property measurement system (MPMS-XL5). The $M(H)$ curves are measured at interval $\Delta T$ = 1 K. The applied magnetic field ($H_a$) has been corrected for the internal field as $H = H_a - NM$, where $M$ is the measured magnetization and $N$ is the demagnetization factor. The corrected $H$ was used for the analysis of critical behavior. The measurement of the M$\mathrm{\ddot{o}}$ssbauer effect in $\mathrm{Fe_{3-x}GeTe_{2}}$ crushed single-crystals was performed in transmission geometry using a $\mathrm{^{57}Co(Rh)}$ source at room temperature ($T = 300$ K). The Wissel spectrometer was calibrated by the spectra of natural iron foil, so the isomer shift values ($\mathit{\delta}$) are in reference to metallic alpha iron ($\mathit{\delta}$ = 0).

\section{RESULTS AND DISCUSSIONS}

Figure 1(a) shows the crystal structure of Fe$_{3-x}$GeTe$_2$, which contains Fe$_{3-x}$Ge slabs separated by van der Waals gapped Te double layers. The Fe atoms in the unit cell occupy two inequivalent Wyckoff sites Fe1 and Fe2, as illustrated in Fig. 1(b). The Fe1 atoms are situated in a hexagonal net in a layer with only Fe atoms. The Fe2 and Ge atoms are covalently bonded in an adjacent layer. The previous study indicated an Fe2 deficient occupancy of 0.866 but full occupancy of Fe1 as well as Ge and Te sites in single crystals grown via chemical vapor transport (CVT).\cite{Zhu} By contrast, our EDX result gives Fe and Ge deficiencies with a composition of Fe$_{2.64(6)}$Ge$_{0.87(4)}$Te$_2$ in the flux-grown single crystals. Figure 1(c) presents the powder x-ray diffraction (XRD) pattern of Fe$_{3-x}$GeTe$_2$, in which the observed peaks are well fitted with the $P6_3/mmc$ space group. The determined lattice parameters $a = 0.3954(2)$ nm and $c = 1.6372(2)$ nm differ from values of $a = 0.40042(15)$ nm and $c = 1.6282(6)$ nm in CVT-grown single crystals.\cite{Zhu} This indicates that the Fe-deficient sample has smaller $a$ and larger $c$ than the Fe-rich sample, in good agreement with the previous report.\cite{Andrew} Furthermore, in the single crystal 2$\theta$ XRD scan [Fig. 1(d)], only $(00l)$ peaks are detected, indicating the crystal surface is normal to the $c$ axis with the plate-shaped surface parallel to the $ab$ plane.

\begin{figure}
\centerline{\includegraphics[scale=0.34]{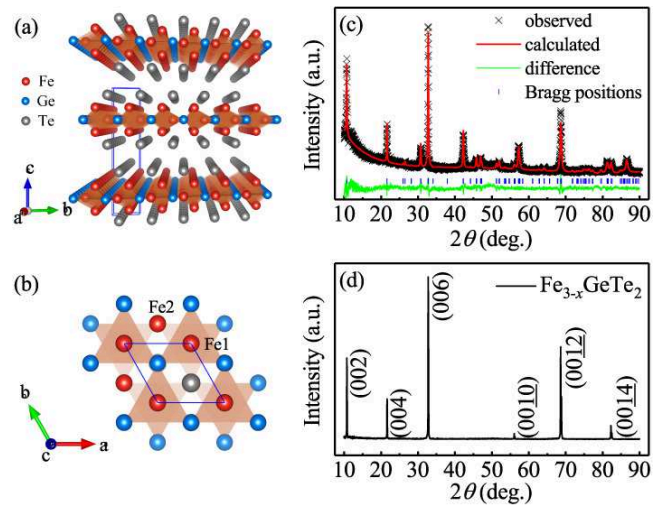}}
\caption{(Color online). Crystal structure of Fe$_{3-x}$GeTe$_2$ from (a) side and (b) top views. The unit cell is enclosed by blue solid lines. Inequivalent Fe sites are labeled as Fe1 and Fe2, respectively. (c) Powder x-ray diffraction (XRD) and (d) single-crystal XRD pattern of Fe$_{3-x}$GeTe$_2$. The vertical tick marks represent Bragg reflections of the $P6_3/mmc$ space group.}
\label{XRD}
\end{figure}

\begin{figure}
\centerline{\includegraphics[scale=1]{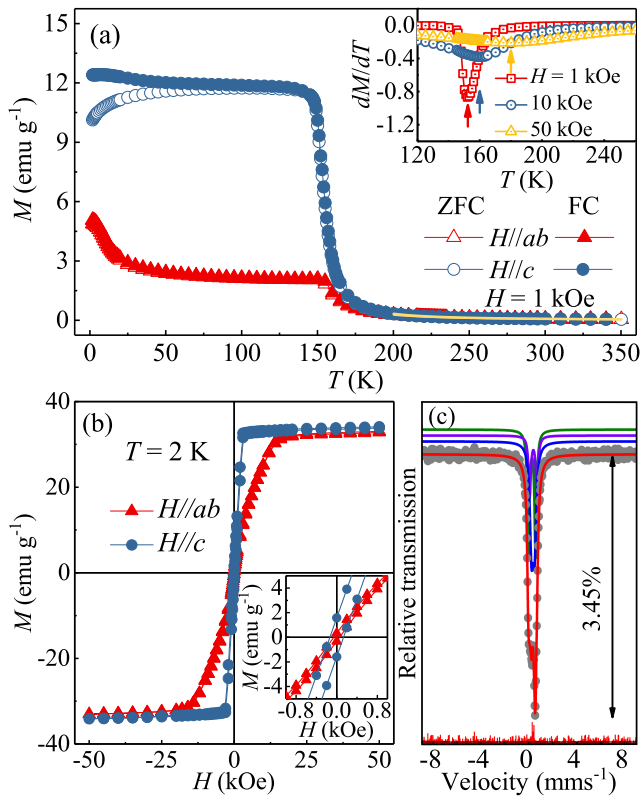}}
\caption{(Color online). (a) Temperature dependence of magnetization for Fe$_{3-x}$GeTe$_2$ measured with the external magnetic field $H$ = 1 kOe applied along the $c$ axis and in the $ab$ plane under zero-field-cooling (ZFC) and field-cooling (FC) modes. The yellow solid lines are fitted by the modified Curie-Weiss law $\chi = \frac{C}{T-\theta}+\chi_0$, where $\chi_0$ is the temperature-independent susceptibility, $C$ is the Curie-Weiss constant, and $\theta$ is the Weiss temperature. Inset: The derivative magnetization $dM/dT$ vs $T$ in different applied fields along the $c$ axis. (b) Field dependence of magnetization for Fe$_{3-x}$GeTe$_2$ measured at $T$ = 2 K. Inset: The magnification in the low field region. (c) M\"{o}ssbauer spectrum at $T = 300$ K of the Fe$_{3-x}$GeTe$_2$ crushed single-crystal. The experimental data are presented by solid circles and the fit is given by the red solid line. Vertical arrow denotes relative position of the lowermost peak with respect to the basal line (relative transmission). The fitted lines of subspectra are plotted above the main spectrum fit: the Fe1 doublet is blue; the Fe1$^{*}$ doublet is violet, and the Fe2 doublet is olive. Error is depicted as the absolute of difference; the largest value is 0.176 \%.}
\label{MTH}
\end{figure}

\begin{figure}
\centerline{\includegraphics[scale=0.9]{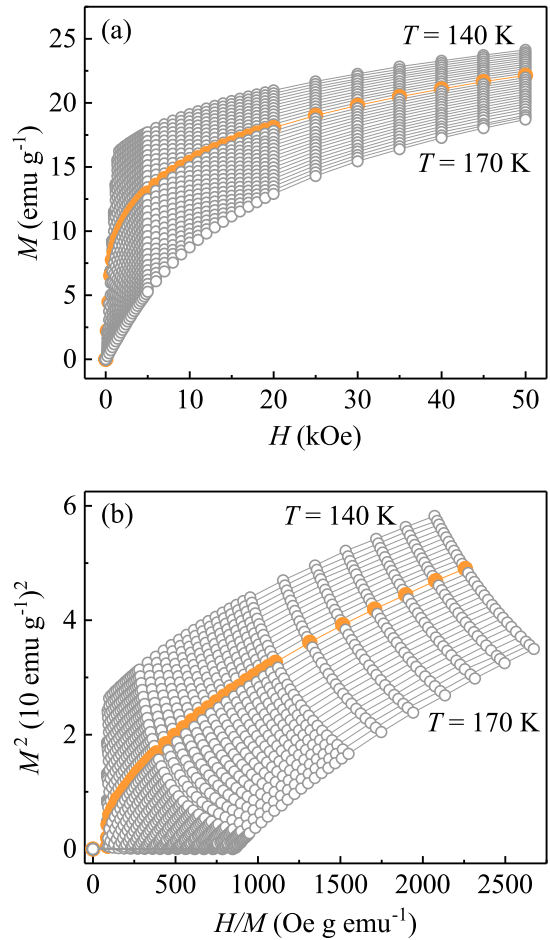}}
\caption{(Color online). (a) Typical initial isothermal magnetization curves measured along the $c$ axis around $T_c$ = 152 K (in an orange symbol and line) for Fe$_{3-x}$GeTe$_2$. (b) Arrott plots of $M^2$ vs $H/M$ around $T_c$ for Fe$_{3-x}$GeTe$_2$.}
\label{Arrot}
\end{figure}

\begin{figure*}
\centerline{\includegraphics[scale=0.9]{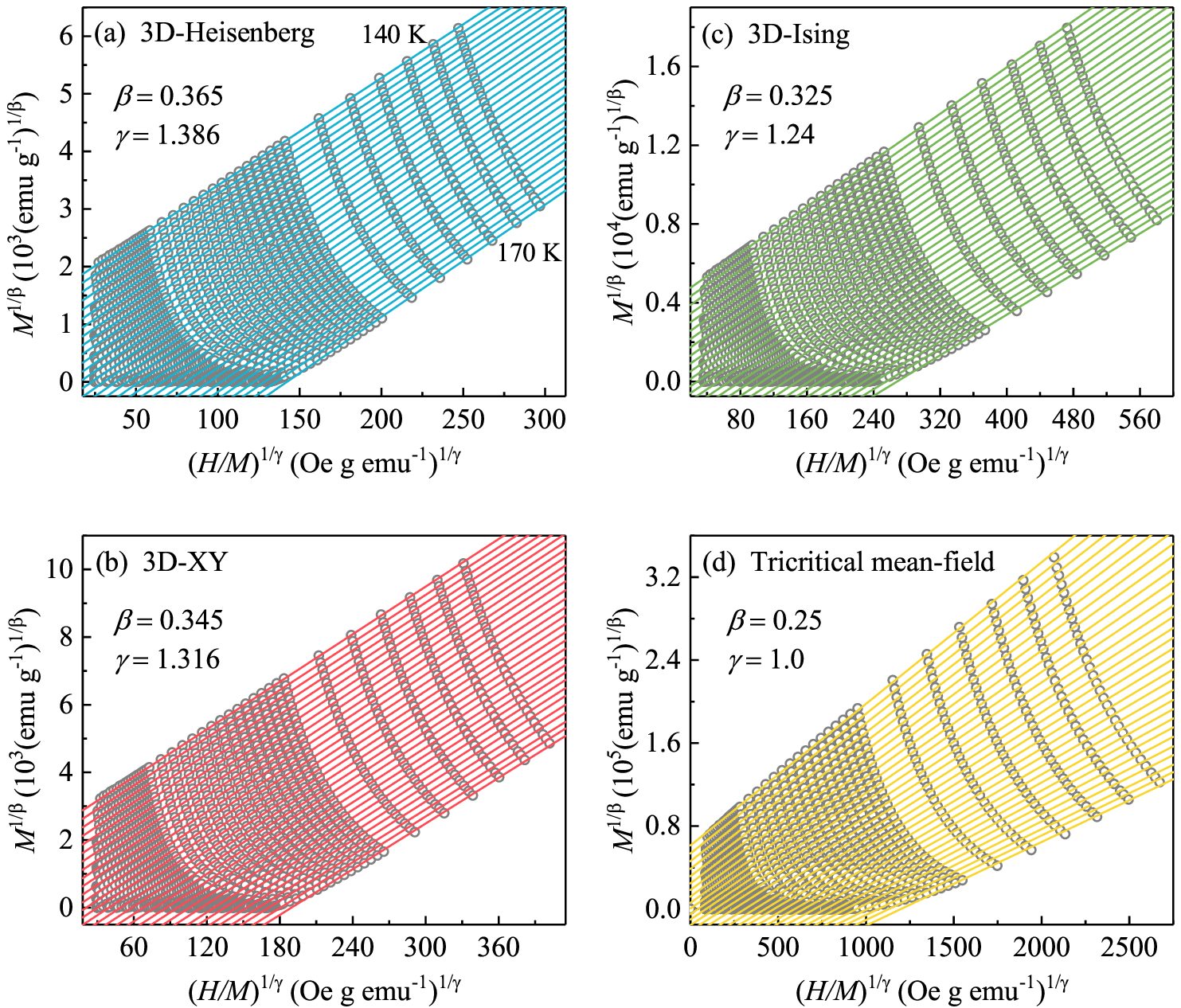}}
\caption{(Color online). The isotherms of $M^{1/\beta}$ vs $(H/M)^{1/\gamma}$ with parameters of (a) 3D Heisenberg model, (b) 3D XY model, (c) 3D Ising model, and (d) tricritical mean-field model. The straight lines are the linear fit of isotherms at different temperatures.}
\label{3D}
\end{figure*}

Figure 2(a) shows the temperature dependence of magnetization $M(T)$ measured under $H$ = 1 kOe applied in the $ab$ plane and parallel to the $c$ axis, respectively. A clear paramagnetic (PM) to ferromagnetic (FM) transition is observed and the apparent anisotropy at low temperatures suggests that the crystallographic $c$ axis is the easy axis. The zero-field-cooling (ZFC) and field-cooling (FC) curves show significant splitting at low temperatures for $H//c$, further indicating strong magnet-crystalline anisotropy in Fe$_{3-x}$GeTe$_2$. Generally, the critical temperature $T_c$ is actually difficult to be determined from $M(T)$ curve because it is usually dependent on the external field. As shown in the inset of Fig. 2(a), $T_c$ increases monotonously with the increase of $H$, which is roughly determined from the minima of the $dM/dT$ curves. The $T_c$ is $\sim 152$ K when $H$ = 1 kOe; it gradually increases to 160 K and 180 K for $H$ = 10 kOe and $H$ = 50 kOe, respectively. In the high temperature region of 200 $\sim$ 350 K, the ZFC curves are well fitted by the modified Curie-Weiss law $\chi = \frac{C}{T-\theta}+\chi_0$, where $\chi_0$ is the temperature-independent susceptibility, $C$ is the Curie-Weiss constant, and $\theta$ is the Weiss temperature. The Weiss temperatures obtained from the fitting are $\theta_{ab} = 157(1)$ K and $\theta_c = 164(1)$ K for $H // ab$ and $H // c$, respectively, the positive value confirming the FM interaction among Fe atoms. The effective moment $\mu_{\textrm{eff}}$ = 4.21(2) $\mu_B/$Fe obtained from $H // ab$ data is identical to $\mu_{\textrm{eff}}$ = 4.19(5) $\mu_B/$Fe from $H//c$ data, indicating a nearly isotropic paramagnetic behavior at high temperatures. Figure 2(b) displays the isothermal magnetization measured at $T$ = 2 K. The saturation field $H_s \approx 3 $ kOe for $H//c$ is much smaller than $H_s \approx 18 $ kOe for $H//ab$, confirming the easy axis is the $c$-axis. The saturation moment at $T$ = 2 K is $M_s \approx$ 1.00(1) $\mu_B/$Fe for $H//ab$ and $M_s \approx$ 1.03(1) $\mu_B/$Fe for $H//c$, respectively. The inset in Fig. 2(b) shows the $M(H)$ in the low field region and little hysteresis with the coercive forces $H_{ab} = 52$ Oe for $H // ab$ and $H_c = 138$ Oe for $H // c$, respectively. All these results are in good agreement with the previous report.\cite{Andrew} Then we calculated the Rhodes-Wohlfarth ratio (RWR) for Fe$_{3-x}$GeTe$_2$, which is defined as $P_c/P_s$ with $P_c$ obtained from the effective moment $P_c(P_c+2) = P_{eff}^2$ and $P_s$ is the saturation moment obtained in the ordered state.\cite{Wohlfarth, Moriya} RWR is 1 for a localized system and is larger in an itinerant system. Here we obtain relatively large values of RWR = 3.33 with $H//ab$ and RWR = 3.21 with $H // c$ for Fe$_{3-x}$GeTe$_2$, which is somewhat smaller than the value of RWR = 3.8 reported in CVT-grown single crystals without Fe vacancy,\cite{Chen1} indicating a possible weak itinerant character and/or strong spin fluctuations in the ground state.

\begin{table}
\caption{M$\mathrm{\ddot{o}}$ssbauer hyperfine parameters at $T$ = 300 K of Fe$_{3-x}$GeTe$_2$. $\mathit{A}$ is the relative area of subspectrum, $\mathit{\Gamma}$ is the line width, $\mathit{\delta}$ is the measured isomer shift, and $\mathit{\Delta}$ is the quadrupole splitting. The fitting errors are presented in parenthesis. The superscript * denotes distorted local enviroment.}
\label{TabMS}
\begin{ruledtabular}
\begin{tabular}{lllll}
 site & $\mathit{A}$ & $\mathit{\Gamma}$ & $\mathit{\delta}$ & $\mathit{\Delta}$ \\
  & [\%] & [mms$^{-1}$] & [mms$^{-1}$] & [mms$^{-1}$] \\
\hline
 Fe$_{1}$ & 54(5) & 0.31(1) & 0.397(2) & 0.25(1) \\
 Fe$_{1}^{*}$ & 16(5) & 0.22(3) & 0.43(1) & 0.59(2) \\
 Fe$_{2}$ & 30(4) & 0.21(1) & 0.309(5) & 0.586(6) \\
\end{tabular}
\end{ruledtabular}
\end{table}

The M\"{o}ssbauer spectrum [Fig. 2(c)] has been examined by WinNormos-Site software package based on the least squares method.\cite{MS1} The fit goodness value is 1.032. The spectrum can be well fitted by three paramagnetic doublets with small discrepancy with the measured spectrum. The discrepancy arises due to choice of the fitting model with equal Lorentz lines. Also, there could be some texture due to residual preferential orientation for the incident angle of $\gamma$-rays in crushed crystals so that the ratio of doublet line areas is different from 1. Hyperfine parameters give insight into first coordination sphere (ICS), i.e. local environment of Fe atoms. ICS for Fe1 consists from Fe1, 3Fe2, 3Ge and 3Te, whereas for Fe2 ICS consists from 3Ge, 2Te and 6Fe1. Fe vacancies prefer Fe2 positions (VFe2) and that contributes to distorted ICS in some Fe1 atoms (Fe1$^{*}$), i.e. will create different local symmetry around such atomic sites. Since the ratio of Fe atoms Fe1 : Fe2 = 2 : 1, we ascribe the largest doublet 54(5)\% (Table I) to Fe1 atomic positions, in agreement with previous results.\cite{Yu} It should be noted that the lattice parameters depend on the number and distribution of vacancies and that electric field gradient (EFG) tensor components are very sensitive on the local atomic bonds of Fe atoms. This is the origin of different quadrupole splitting ($\Delta$) (Table I). For Fe1$^{*}$ atoms we expect larger values of $\Delta$ due to additional asymmetry of EFG ($\eta$) that stems from the local symmetry distortion (Table I).\cite{MS3} Small change of isomer shift (Table I) suggests that vacancies cause a redistribution of valence electrons. In contrast to Fe1, the shortest chemical bonds for Fe2 are with Ge atoms and are covalent.\cite{MS4} This brings different chemical shift (electric monopol interaction E$_{0}$) when compared to Fe1. This and second order Doppler shift (also different for different local environment of Fe1, Fe1$^{*}$ and Fe2) contribute to isomer shift $\delta$. Hence, the doublet with $\delta$ = 0.309(5) mm$^{-1}$ is ascribed to Fe2 in 2c position with -6m2 point group symmetry.\cite{Deiseroth} This is higher symmetry when compared to 3m of Fe1 in 4e position. Measured $\Delta$ (Table I) suggest that charge density around Fe2 is more anisotropic when compared to Fe1. Ratio Fe1$^{*}$ : Fe1 assuming similar recoilless factors for Fe1 and Fe1$^{*}$ points to vacancies on Fe2 atomic site that corresponds to about 2.2\% distorted ICS of Fe2. Doublets or distribution of quadrupole splitting were not detected. Based on the linewidth ($\Gamma$, Table I), Fe1 ICS is less ordered when compared to Fe1$^{*}$ and Fe2. This could be due to larger distances of Fe2 vacancies from other ICS so that Fe2 vacancies have negligible influence on Fe1 ICS. The other possibility could be that Fe2 vacancies take preferential positions in ICS of Fe1 with higher local symmetry, thereby decreasing $\eta$ and $\Delta$ parameters. Theoretical calculations of hyperfine parameters could shed more light on this.

As is well known, the critical behavior of a second-order transition can be characterized in detail by a series of interrelated critical exponents.\cite{Stanley} In the vicinity of a second-order phase transition, the divergence of correlation length $\xi = \xi_0 |(T-T_c)/T_c|^{-\nu}$ leads to universal scaling laws for the spontaneous magnetization $M_s$ and the inverse initial magnetic susceptibility $\chi_0^{-1}$. The spontaneous magnetization $M_s$ below $T_c$, the inverse initial susceptibility $\chi_0^{-1}$ above $T_c$, and the measured magnetization $M(H)$ at $T_c$ are characterized by a set of critical exponents $\beta$, $\gamma$, and $\delta$. The mathematical definitions of these exponents from magnetization are:
\begin{equation}
M_s (T) = M_0(-\varepsilon)^\beta, \varepsilon < 0, T < T_c,
\end{equation}
\begin{equation}
\chi_0^{-1} (T) = (h_0/m_0)\varepsilon^\gamma, \varepsilon > 0, T > T_c,
\end{equation}
\begin{equation}
M = DH^{1/\delta}, \varepsilon = 0, T = T_c,
\end{equation}
where $\varepsilon = (T-T_c)/T_c$ is the reduced temperature, and $M_0$, $h_0/m_0$, and $D$ are the critical amplitudes.\cite{Fisher} The magnetic equation of state is a relationship among the variables $M(H,\varepsilon)$, $H$, and $T$. Using the scaling hypothesis this can be expressed as
\begin{equation}
M(H,\varepsilon) = \varepsilon^\beta f_\pm(H/\varepsilon^{\beta+\gamma}),
\end{equation}
where $f_+$ for $T>T_c$ and $f_-$ for $T<T_c$, respectively, are the regular functions. In terms of renormalized magnetization $m\equiv\varepsilon^{-\beta}M(H,\varepsilon)$ and renormalized field $h\equiv\varepsilon^{-(\beta+\gamma)}H$, Eq.(4) can be written as
\begin{equation}
m = f_\pm(h),
\end{equation}
which implies that for true scaling relations and the right choice of $\beta$, $\gamma$, and $\delta$ values, scaled $m$
and $h$ will fall on two universal curves: one above $T_c$ and another below $T_c$. This is an important criterion for the critical regime.

\begin{figure}
\centerline{\includegraphics[scale=1]{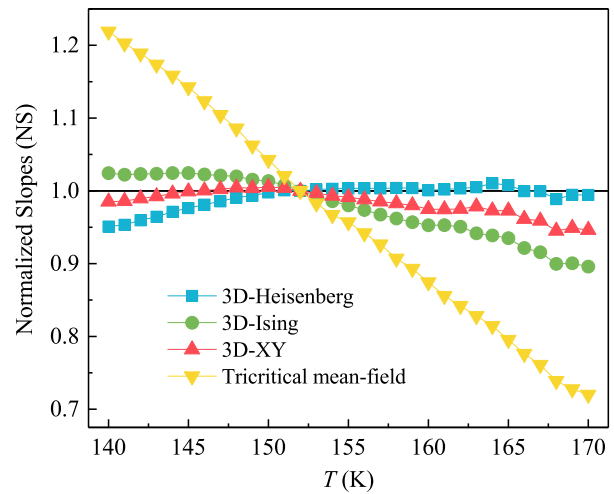}}
\caption{(Color online). Temperature dependence of the normalized slopes $NS = S(T)/S(T_c)$.}
\label{NS}
\end{figure}

In order to clarify the nature of the PM-FM transition in Fe$_{3-x}$GeTe$_2$, we measured the isothermal $M(H)$ in the temperature range from $T$ = 140 K to $T$ = 170 K, as shown in Fig. 3(a). Generally, the conventional method to determine the critical exponents and critical temperature involves the use of an Arrott plot.\cite{Arrott1} The Arrott plot assumes critical exponents following the mean-field theory with $\beta$ = 0.5 and $\gamma$ = 1.0.\cite{Arrott1} According to this method, isotherms plotted in the form of $M^2$ vs $H/M$ constitute a set of parallel straight lines, and the isotherm at the critical temperature $T_c$ should pass through the origin. At the same time, it directly gives $\chi_0^{-1}(T)$ and $M_s(T)$ as the intercepts on the $H/M$ axis and positive $M^2$ axis, respectively. Figure 3(b) shows the Arrott plot of Fe$_{3-x}$GeTe$_2$. All the curves in this plot show nonlinear behavior having a downward curvature even in high fields. This suggests that the framework of Landau mean-field model is not applicable to Fe$_{3-x}$GeTe$_2$. According to Banerjee$^\prime$s criterion,\cite{Banerjee} one can estimate the order of the magnetic transition through the slope of the straight line: A negative slope corresponds to the first-order transition while positive corresponds to the second order. Therefore, the concave downward curvature clearly indicates that the PM-FM transition in Fe$_{3-x}$GeTe$_2$ is a second-order one.

\begin{figure}
\centerline{\includegraphics[scale=1]{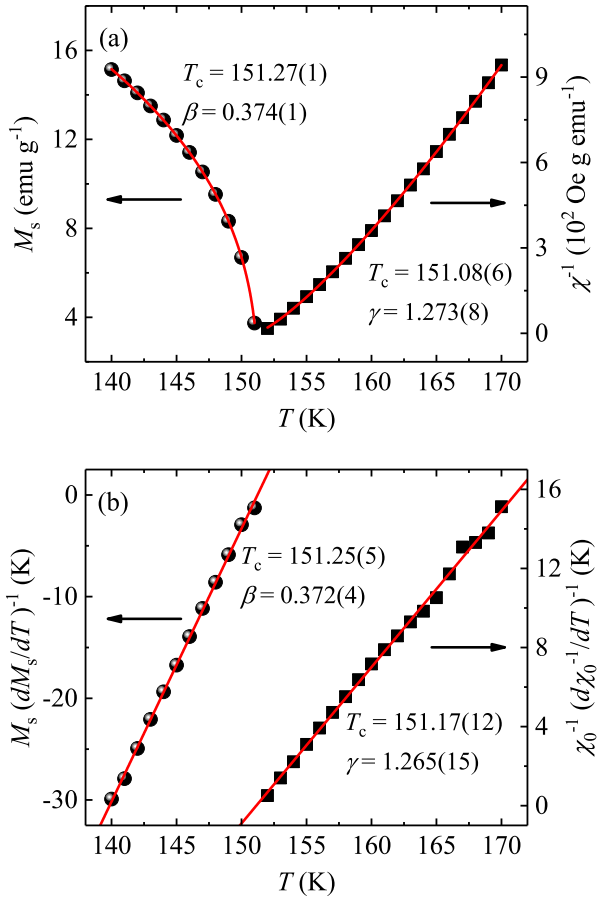}}
\caption{(Color online). (a) Temperature dependence of the spontaneous magnetization $M_s$ (left) and the inverse initial susceptibility $\chi_0^{-1}$ (right) with solid fitting curves for Fe$_{3-x}$GeTe$_2$. (b) Kouvel-Fisher plots of $M_s(dM_s/dT)^{-1}$ (left) and $\chi_0^{-1}(d\chi_0^{-1}/dT)^{-1}$ (right) with solid fitting curves for Fe$_{3-x}$GeTe$_2$.}
\label{KF}
\end{figure}

A modified Arrott plot of $M^{1/\beta}$ vs $(H/M)^{1/\gamma}$ could be employed.\cite{Arrott2} The modified Arrott plot is given by the Arrot-Noaks equation of state
\begin{equation}
(H/M)^{1/\gamma} = a\varepsilon+bM^{1/\beta},
\end{equation}
where $\varepsilon = (T-T_c)/T_c$ is the reduced temperature, and $a$ and $b$ are constants. Four kinds of possible exponents belonging to a 3D Heisenberg model ($\beta = 0.365, \gamma = 1.386$), 3D XY model ($\beta = 0.345, \gamma = 1.316$), 3D Ising model ($\beta = 0.325, \gamma = 1.24$), and tricritical mean-field model ($\beta = 0.25, \gamma = 1.0$) are used to construct the modified Arrott plots,\cite{Zhang3} as shown in Fig. 4. As we can see, all these four constructions exhibit quasi straight lines in the high field region. Apparently, the lines in Fig. 4(d) are not parallel to each other, indicating that the tricritical mean-field model is not satisfied. However, all lines in Figs. 4(a)-(c) are almost parallel to each other. To determine an appropriate model, the modified Arrott plots should be a series of parallel lines in the high field region with the same slope, where the slope is defined as $S(T) = dM^{1/\beta}/d(H/M)^{1/\gamma}$. The normalized slope ($NS$) is defined as $NS = S(T)/S(T_c)$, which enables us to identify the most suitable model by comparing the $NS$ with the ideal value of 1. Plot of $NS$ vs $T$ for the four different models is shown in Fig. 5. One can see that the $NS$ of 3D Heisenberg model almost equals to $NS = 1$ above $T_c$, in accordance with the nearly isotropic magnetic character at high temperatures [Fig. 2(a)], while that of 3D XY model is the best below $T_c$, indicating an enhancement of the anisotropic interaction (spin fluctuations) on cooling through the transition point.

To obtain the precise critical exponents $\beta$ and $\gamma$, a rigorous iterative method has been used.\cite{Pramanik} The linear extrapolation from the high field region to the intercepts with the axis $M^{1/\beta}$ and $(H/M)^{1/\gamma}$ yields reliable values of $M_s(T)$ and $\chi_0^{-1}(T)$. A set of $\beta$ and $\gamma$ can be obtained by fitting the data following the Eqs. (1) and (2). Then the obtained new values of $\beta$ and $\gamma$ are used to reconstruct a new modified Arrott plot. Consequently, new $M_s(T)$ and $\chi_0^{-1}(T)$ are generated from the linear extrapolation from the high field region. Therefore, another set of $\beta$ and $\gamma$ can be generated. This procedure was repeated until the values of $\beta$ and $\gamma$ are stable. From this method, the obtained critical exponents are hardly dependent on the initial parameters, which confirms these critical exponents are reliable and intrinsic. Figure 6(a) presents the final $M_s(T)$ and $\chi_0^{-1}(T)$ with solid fitting curves. The critical exponents $\beta = 0.374(1)$ with $T_c = 151.27(1)$ K and $\gamma = 1.273(8)$ with $T_c = 151.08(6)$ K are obtained. Alternatively, the critical exponents can be determined by the Kouvel-Fisher (KF) method,\cite{Kouvel}
\begin{equation}
\frac{M_s(T)}{dM_s(T)/dT} = \frac{T-T_c}{\beta},
\end{equation}
\begin{equation}
\frac{\chi_0^{-1}(T)}{d\chi_0^{-1}(T)/dT} = \frac{T-T_c}{\gamma}.
\end{equation}
According to this method, $M_s(T)/[dM_s(T)/dT]$ and $\chi_0^{-1}(T)/[d\chi_0^{-1}(T)/dT]$ are as linear functions of temperature with slopes of $1/\beta$ and $1/\gamma$, respectively. As shown in Fig. 6(b), the linear fits give $\beta = 0.372(4)$ with $T_c = 151.25(5)$ K and $\gamma = 1.265(15)$ with $T_c = 151.17(12)$ K, respectively.

\begin{figure}
\centerline{\includegraphics[scale=1]{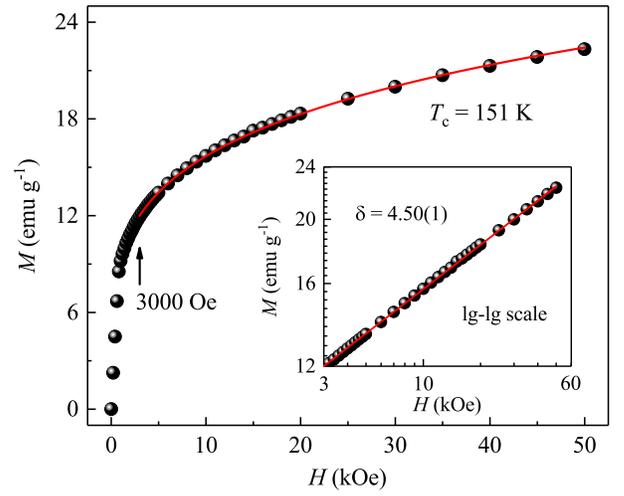}}
\caption{(Color online). Isotherm $M$ vs $H$ plot collected at $T_c$ = 151 K for Fe$_{3-x}$GeTe$_2$. Inset: The same plot in log-log scale with a solid fitting curve.}
\label{MH}
\end{figure}

\begin{figure}
\centerline{\includegraphics[scale=1]{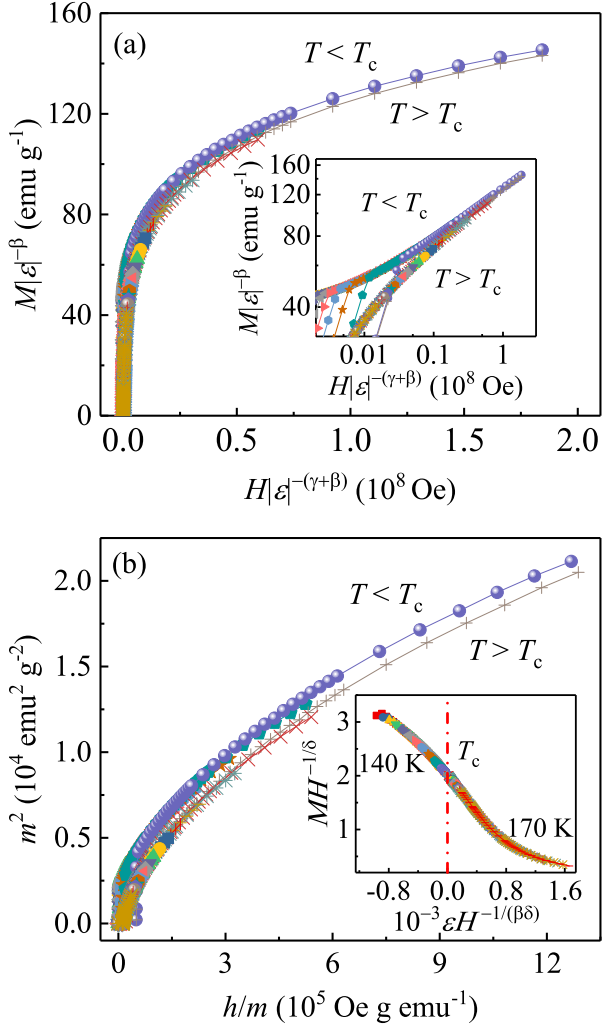}}
\caption{(Color online). (a) Scaling plots of renormalized magnetization $m$ vs renormalized field $h$ below and above $T_c$ for Fe$_{3-x}$GeTe$_2$. Inset: The same plots in log-log scale. (b) The renormalized magnetization and field replotted in the form of $m^2$ vs $h/m$ for Fe$_{3-x}$GeTe$_2$. Inset: The rescaling of the $M(H)$ curves by $MH^{-1/\delta}$ vs $\varepsilon H^{-1/(\beta\delta)}$.}
\label{magnetism}
\end{figure}

The third exponent $\delta$ can be calculated from Widom scaling law,
\begin{equation}
\delta = 1+\frac{\gamma}{\beta}.
\end{equation}
Using the $\beta$ and $\gamma$ values determined from the modified Arrott plot and Kouvel-Fisher plot, we obtain $\delta$ = 4.404(12) and $\delta$ = 4.401(6), respectively. Isothermal magnetization $M(H)$ at a critical temperature $T_c$ = 151 K is shown in Fig. 7, with the inset plotted on a lg-lg scale. According to Eq. (3), the critical exponent $\delta = 4.50(1)$ can be deduced, which is very close to the values obtained from the modified Arrott plot and Kouvel-Fisher plot. Therefore, the critical exponents $\beta$, $\gamma$, $\delta$, and $T_c$ obtained in the present study are self-consistent and accurately estimated within experimental precision. We note that these values are close to what was found in stoichiometric Fe$_{3}$GeTe$_{2}$,\cite{LiuB} suggesting that vacancies have little influence on critical regime and dimensionality of magnetic interactions, in contrast to $T_{c}$ value.

The reliability of the obtained critical exponents and $T_c$ can also be verified by a scaling analysis. Following Eq. (5), scaled $m$ vs scaled $h$ has been plotted in Fig. 8(a), along with the same plot on a log-log scale in the inset of Fig. 8(a). It is rather significant that all the data collapse into two separate branches: one below $T_c$ and another above $T_c$. The reliability of the exponents and $T_c$ has been further ensured with a more rigorous method by plotting $m^2$ vs $h/m$, as shown in Fig. 8(b), where all data also fall on two independent branches. This clearly indicates that the interactions get properly renormalized in a critical regime following the scaling equation of state. In addition, the scaling equation of state takes another form,
\begin{equation}
\frac{H}{M^\delta} = k(\frac{\varepsilon}{H^{1/\beta}}),
\end{equation}
where $k(x)$ is the scaling function. Based on Eq. (10), all experimental curves will collapse into a single curve. The inset of Fig. 8(b) shows the $MH^{-1/\delta}$ vs $\varepsilon H^{-1/(\beta\delta)}$ for Fe$_{3-x}$GeTe$_2$, where the experimental data collapse into a single curve, and $T_c$ locates at the zero point of the horizontal axis. The well-rescaled curves further confirm the reliability of the obtained critical exponents.\cite{LiuB}

\begin{table*}
\caption{\label{tab1}Comparison of critical exponents of Fe$_{3-x}$GeTe$_2$ with different theoretical models.}
\begin{ruledtabular}
\begin{tabular}{lllllll}
  Composition & Reference & Technique & $T_c$ & $\beta$ & $\gamma$ & $\delta$ \\
  \hline
  Fe$_{3-x}$GeTe$_2$ & This work & Modified Arrott plot & 151.27(1) & 0.374(1) & 1.273(8) & 4.404(12) \\
  & This work & Kouvel-Fisher plot & 151.25(5) & 0.372(4) & 1.265(15) & 4.401(6) \\
  & This work &Critical isotherm  &   &   &   & 4.50(1) \\
  3D Heisenberg & 28 & Theory & & 0.365 & 1.386 & 4.8 \\
  3D XY & 28 & Theory & & 0.345 & 1.316 & 4.81 \\
  3D Ising & 28 & Theory & & 0.325 & 1.24 & 4.82 \\
  Tricritical mean field & 36 & Theory & & 0.25 & 1.0 & 5.0
\end{tabular}
\end{ruledtabular}
\end{table*}

The obtained critical exponents of Fe$_{3-x}$GeTe$_2$, as well as those of different theoretical models, are listed in Table II for comparison. Taroni \emph{et al.} have accomplished a comprehensive investigation of critical exponents for 2D magnets with a conclusion that the critical exponent $\beta$ for a 2D magnet should be within a window $\sim$ $0.1 \leq \beta \leq 0.25$.\cite{Taroni} That is to say, the critical exponents of Fe$_{3-x}$GeTe$_2$ exhibit apparent 3D critical phenomenon. One can see that the critical exponent $\beta$ of Fe$_{3-x}$GeTe$_2$ is close to that of 3D Heisenberg model. While $\gamma$ approaches to that of 3D XY and/or 3D Ising model, which might be the origin of large magnetocrystalline anisotropy in the ground state of Fe$_{3-x}$GeTe$_2$. Then it is important to understand the nature as well as the range of interaction in this material. As we know, for a homogeneous magnet, the universality class of the magnetic phase transition depending on the exchange distance $J(r)$. Fisher \emph{et al.} theoretically treated this kind of magnetic ordering as an attractive interaction of spins, where a renormalization group theory analysis suggests the interaction decays with distance $r$ as
\begin{equation}
J(r) \approx r^{-(3+\sigma)},
\end{equation}
where $\sigma$ is a positive constant.\cite{Fisher1972} Moreover, the susceptibility exponent $\gamma$ is predicted as,
\begin{multline}
\gamma = 1+\frac{4}{d}(\frac{n+2}{n+8})\Delta\sigma+\frac{8(n+2)(n-4)}{d^2(n+8)^2}\\\times[1+\frac{2G(\frac{d}{2})(7n+20)}{(n-4)(n+8)}]\Delta\sigma^2,
\end{multline}
where $\Delta\sigma = (\sigma-\frac{d}{2})$ and $G(\frac{d}{2})=3-\frac{1}{4}(\frac{d}{2})^2$, $n$ is the spin dimensionality. When $\sigma > 2$, the Heisenberg model is valid for the 3D isotropic magnet, where $J(r)$ decreases faster than $r^{-5}$. When $\sigma \leq 3/2$, the mean-field model is satisfied, expecting that $J(r)$ decreases slower than $r^{-4.5}$. From Eq. (12) it is found that \{$d:n$\} = \{3 : 3\} and $\sigma = 1.89$ give the exponents ($\beta = 0.391$, $\gamma = 1.332$, and $\delta = 4.407$), which are mostly close to our experimentally observed values (Table II). Moreover, we obtain the correlation length critical exponent $\nu$ = 0.705 ($\nu = \gamma/\sigma$, $\xi = \xi_0 |(T-T_c)/T_c|^{-\nu}$), and $\alpha$ = -0.115 ($\alpha= 2 - \nu d$), which is close to the theoretical value for 3D Heisenberg model [$\alpha$ = -0.115(9)].\cite{LeGuillou, Fisher1974} This calculation suggests that the spin interaction in Fe$_{3-x}$GeTe$_2$ is close to the 3D Heisenberg (\{$d:n$\} = \{3 : 3\}) type coupled with a long-range ($\sigma = 1.89$) interaction.

\section{CONCLUSIONS}

In summary, we have made a comprehensive study on the critical phenomenon at the PM-FM phase transition in the van der Waals bonded ferromagnet Fe$_{3-x}$GeTe$_2$. This transition is identified to be second order in nature. The critical exponents $\beta$, $\gamma$, and $\delta$ estimated from various techniques match reasonably well and follow the scaling equation, confirming that the obtained exponents are unambiguous and intrinsic to the material. The determined exponents are close to a 3D Heisenberg ($d = 3, n = 3$) spins coupled with a long-range interaction between spins decay as $J(r)\approx r^{-(3+\sigma)}$ with $\sigma=1.89$. Furthermore, with the rapid development in the field of 2D materials, we expect our experimental work to stimulate broad interests in reducing bulk Fe$_{3-x}$GeTe$_2$ to monolayer sheets and possible spintronic application.

\section*{Acknowledgments}
We thank John Warren for help with the scanning electron microscopy (SEM) measurements. This work was supported by the U.S. DOE-BES, Division of Materials Science and Engineering, under Contract No. DE-SC0012704 (BNL) and by the grant No. 171001 by the Ministry of Education, Science and Technological Development of the Republic of Serbia.

\end{document}